\documentclass[prd,aps,twocolumn,preprintnumbers,10pt]{revtex4}
\usepackage{amssymb,amsmath,amsfonts,amsbsy,epsfig}

\newcommand{\be}{\begin{equation}}
\newcommand{\ee}{\end{equation}}
\newcommand{\bea}{\begin{eqnarray}}
\newcommand{\eea}{\end{eqnarray}}
\newcommand{\nn}{\nonumber}

\def\p{{\bf p}}
\def\q{{\bf q}}
\def\k{{\bf k}}
\def\x{{\bf x}}

\begin{document}

\title{On the dynamics of fluctuations in time crystals
}
\date{\today}
\author{
Esteban Castillo$^{a}$, Benjamin Koch$^{b}$, and Gonzalo Palma$^{c}$
}
\affiliation{
\vspace{5pt}
$^{a}$Instituto de Astrof\'{i}sica, Pontificia Universidad Cat\'{o}lica de Chile, \mbox{Avenida Vicu\~na Mackenna 4860, Santiago, Chile}
\\
$^{b}$Instituto de F\'{i}sica, Pontificia Universidad Cat\'{o}lica de Chile, \mbox{Avenida Vicu\~na Mackenna 4860, Santiago, Chile}
\\
$^{c}$Departamento de F\'{i}sica, FCFM, Universidad de Chile, \mbox{Blanco Encalada 2008, Santiago, Chile}
}

\begin{abstract} 

We study the evolution of quantum fluctuations in systems known as \emph{time crystals},
hypothetical systems for which the lowest energy state performs a periodic motion.
We first discuss some general properties shared by time crystals, and deduce the effective field theory parametrizing the evolution of their fluctuations. We show that these fluctuations fall into categories analogous to acoustic and optical phonons, encountered in conventional crystals. The acoustic phonons correspond to gapless Goldstone boson modes parametrizing the broken time translation invariance of the crystal, whereas the optical phonons are identified with modes perpendicular to the broken symmetry of the system, which generically remain gapped. We study how these two modes decay and interact together, and discuss some observable features that could be tested in experimental realizations of time crystals.
\end{abstract}

\maketitle 

\section{Introduction}

The term ``time crystal" was recently coined by Wilczek and Shapere~\cite{wilczek12, shapere12} in order to characterize a class of systems for which time translation symmetry is spontaneously broken in such a way that the periodic motion of the background constitutes its lowest energy state. While the prospects of their experimental realization has become the subject of some debate~\cite{coleman2013, li12, Schaden:2012dp, Bruno2013, Bruno:2013rdc, 2012arXiv1212.6959L, Wilczek:2013uca, Yoshii:2014fwa, Nozieres}, their theoretical conception has given rise to a fruitful framework to study the properties of relativistic systems with broken Lorentz invariance~\cite{nicolis11, nicolis12, Zhao:2012dp, Nicolis:2013sga, Watanabe:2013uya}. In this note, we study this class of systems aiming to identify generic properties that could be tested in the future, once their experimental realization is better understood.

To be concrete, time crystals are systems for which time translation invariance is broken with the help of an internal compact symmetry in the following way: If $H$ is the Hamiltonian of the full system, and $Q$ the generator of the internal symmetry, it is then possible to have configurations characterized by a non-vanishing parameter $\mu$ such that the following linear combination
\be
\mathcal H \equiv H - \mu \, Q ,  \label{unbroken-H}
\ee
remains unbroken~\cite{nicolis12}. Then, there exists an eigenstate $| \mu \rangle$ that is the state of lowest eigenvalue of $\mathcal H$, for which we can choose $\mathcal H | \mu \rangle = 0$. If the symmetry in question is part of a larger symmetry group, with an algebra spanned by a set of generators $T_a$, then (\ref{unbroken-H}) implies the existence of solutions breaking time translation invariance along a specific direction of the algebra, parametrized by a constant vector $t^a$, such that $Q =  t^a T_a$.

For the state $| \mu \rangle$ the Hamiltonian acts as the symmetry generator $H | \mu \rangle = \mu \, Q | \mu \rangle $, implying that the background is probing the internal symmetry by performing a periodic motion. This may be seen as follows: First, consider that the field content of the theory includes a set of fields $\hat \phi^i(t,\x)$ transforming under a given representation of the symmetry group as $[\hat \phi^i, Q] =  i Q^p{i}_{j} \, \hat \phi^j $, where $Q^{i}_{j} \equiv t^{a} [T_a]^{i}_{j}$. Then, if we define $\phi_0^i (t) \equiv \langle \mu | \hat \phi^i | \mu \rangle$, it follows from $i \dot \phi_0^i (t) \equiv \langle \mu | [ \hat \phi^i , H]  | \mu \rangle$ that $i \dot \phi_0^i (t) =  \mu \, \langle \mu | [ \hat \phi^i , Q]  | \mu \rangle =  \mu \,  t^{a} [T_a]^{i}_{j}  \, \phi_0^j(t)$, which constitutes an equation of motion respected by the background fields (we are working with natural units $\hbar = 1$ and $c=1$). The solution to this equation is
\be
\phi_0^i (t) = \exp\{ \mu \, t \, Q^{i}_{j} \} \phi_0^j(0),
\ee
from where we see that, indeed, the background probes the direction $t^a$ of the internal symmetry.

\section{EFT for time crystals}

To deduce the effective field theory (EFT) for time crystals, it is helpful to introduce a Goldstone boson field $\pi(t,x)$ as the fluctuation along the broken time translation symmetry~\cite{Goldstone:1961eq} defined through the following time reparametrization of the background fields: $t \to \bar t = t + \pi (t,x)$. The introduction of the Goldstone boson in this way allows us to keep track of the fluctuations along the trajectory followed by the background fields $\phi_0^i$. This is achieved by defining $\hat \psi^i (t,\x) \equiv \phi_0^i (t + \pi(t,\x))$
 and demanding that $\pi(t,\x)$ fluctuates in such a way that 
\be
[\hat \psi^i , H] = i \partial_t \hat \psi^i . \label{H-psi}
\ee
This requirement identifies $\hat \psi^i(t,\x)$ with the fluctuations $\hat \phi^i(t,\x)$ projected along the trajectory. Explicitly, we have:
\be
\hat \psi_0^i (t, \x) \equiv \exp\{ \mu \,(  t + \pi(t, \x)  ) \, Q^{i}_{j} \} \phi_0^j(0)  . \label{psi-pi}
\ee
Notice that (\ref{H-psi}) together with (\ref{psi-pi}) implies that $[\pi , H] = i + i \dot \pi$. On the other hand, because $[\hat \psi^i , Q] = i Q^{i}_{j} \hat \psi^j$, it follows that $[\pi, Q] = i / \mu$. Putting these two results together allows one to deduce that $ \mathcal H$ is nothing but the generator of time translations for the fluctuations about the time dependent background fields $\phi_0^i(t)$:
\be
[\pi , \mathcal H] = i \dot \pi. 
\ee
Now, the action for the Goldstone boson $\pi$ is derived from the full action containing the $\hat \phi^i(t, \x)$ fields, which is invariant under the symmetry group containing $Q$ as one of its generators. Thus, even if $Q$ is spontaneously broken by the background solutions $\phi_0^i(t)$, the action for $\pi$ must be invariant under the transformation:
\bea
\hat \psi_0^i (t, \x) \to {\hat \psi_0^{i}}{'} (t, \x) =  \exp\{ \epsilon \, \mu \, Q^{i}_{j} \} \hat \psi_0^j (t, \x) .
\eea
This transformation is equivalent to a shift along the trajectory in field space, or equivalently, to a shift of the $\pi$ fluctuation. Thus, we conclude that the action for the Goldstone boson $\pi$ must be invariant under shifts:
\be
\pi (t,\x)\to \pi'(t,\x) = \pi(t,\x) + \epsilon.
\ee
It follows that background quantities appearing in the EFT action for $\pi(t,\x)$ must be time independent and, therefore, that $\pi(t,\x)$ will appear in it only through gradients of the form $\partial_\mu (t + \pi)$. Moreover, if we assume that the full original theory is invariant under Lorentz transformations, $\pi(t,\x)$ will appear in its action through the following combination $\Pi= \eta^{\mu \nu} \partial_{\mu} (t + \pi) \partial_{\nu} (t + \pi)$, or written more conveniently:
\be
 \Pi = \dot \pi  + \frac{1}{2} \left[  \dot \pi^2 - (\nabla \pi)^2 \right] . \label{basic-brick}
\ee
Thus, the EFT action for $\pi$ may be written as an expansion in powers of $\Pi$ of the following form
\be
S = \Omega^4  \sum_{n=1}^{\infty} \int_{x}    \alpha_n \left( \dot \pi + \frac{1}{2}\dot \pi^2 - \frac{1}{2} (\nabla \pi)^2 \right)^{n}   , \label{EFT-pi-0}
\ee
where $\int_x$ stands for $\int d^4 x$. Here, $\alpha_{n}$ are 
dimensionless constants parametrizing the perturbative expansion, and $\Omega$ is a mass scale introduced to keep $S$ dimensionless. To gain familiarity with this theory, we may write it up to cubic order. In this case we obtain
\bea
S &=& \frac{\Omega^4}{2 c_s^2} \int_x \bigg[  \dot \pi^2 - c_s^2 (\nabla \pi)^2  \nn \\
&& + \left(1- c_s^2 \right)  \dot \pi \left( \dot \pi^2 - (\nabla \pi)^2 \right)  + 2 \alpha_3 \dot \pi^3   \bigg] , \label{EFT-pi}
\eea
where $c_s$ corresponds to the speed of sound at which Goldstone boson fluctuations propagate, and is given in terms of $\alpha_2$ as $1/c_s^2 = 1 + 2 \alpha_2$. Notice that we have set $\alpha_{1} = 1$ to normalize the expansion, and have thrown away a total time derivative.
After this identification, the canonical Goldstone boson $\pi_c(t, \x)$ may be obtained in terms of $\pi(t, \x)$ by the rescaling
$\pi_c \equiv  \Omega^2  c_s^{-1} \, \pi$.
It is worth noticing that the coefficients $\alpha_n$ contain all the relevant information pertaining the background and, therefore, they depend on quantities that may be tuned to study the behavior of the time crystal. For instance, tuning the value of $\alpha_2$ changes both, the value of the speed of sound and the cubic coupling in front of the term $\dot \pi \left( \dot \pi^2 - (\nabla \pi)^2 \right)$, establishing a non-linear relation between the quadratic part and the cubic part of the action. This general result for time crystals is due to the fact that the action was derived using the basic combination (\ref{basic-brick}), which necessarily induces relations between different orders of the perturbation theory. In fact, this has been one of the salient points of the effective field theory of inflation, recently developed to study the evolution of quantum primordial perturbations produced during the first instances of our universe, where time translation invariance is broken in a similar way to the one discussed here~\cite{cheung07a, senatore09a, Achucarro:2012sm, gwyn12, Cespedes:2013rda, Gwyn:2014doa}. 

We may now consider the inclusion of other fluctuations interacting with the Goldstone boson $\pi$. These could correspond to fields parametrizing fluctuations ortogonal to $\hat \psi^i_0(t,\x)$. To simplify our discussion, let us consider a single field $\sigma(t,\x)$ representing this additional sector. The effective field theory appearing from the addition of $\sigma(t,\x)$ must be such that we recover (\ref{EFT-pi}) whenever we set $\sigma = 0$. Recognizing this allows us to plug $\sigma(t,\x)$ into (\ref{EFT-pi}) in every possible way as long as $\pi =$constant and $\sigma = 0$ are stable solutions of the system~\cite{Achucarro:2012sm}. This EFT would admit, for instance, arbitrary powers of $\partial_\mu \sigma \partial^{\mu} \sigma$ and couplings of the form $\partial^{\mu} \sigma \partial_{\mu}(t + \pi)$. Instead of writing the most general action for $\pi$ and $\sigma$ (a task that would not illuminate our discussion of this class of system) let us restrict ourselves to the case where $\sigma$ is a canonical field coupled to powers of the operator (\ref{basic-brick}). This action has the form
\bea
S  \! &=& \! \frac{1}{2} \int_x  \bigg[ \dot \sigma^2 - (\nabla \sigma)^2 - 2 V(\sigma) +  \nn \\ 
&& + 2 \Omega^4 \sum_{n=1}^{\infty} \bar \alpha_n \left( \sigma/\Omega \right)  \left( \dot \pi + \frac{1}{2}\dot \pi^2 - \frac{1}{2} (\nabla \pi)^2 \right)^{n}    \bigg] , \label{EFT-pi-sigma} \qquad
\eea
where the $\bar \alpha_n$'s are dimensionless functions of $\sigma/ \Omega$. The function $V(\sigma)$ corresponds to a scalar potential with its minimum at $\sigma = 0$.  By expanding this action to cubic order, and writing it in terms of a canonically normalized Goldstone $\pi_c$ boson, it becomes:
\bea
S \!\! &=& \!\! \frac{1}{2} \int_x \bigg[  \dot \pi_c^2 - s^2 (\nabla \pi_c)^2  + \dot \sigma^2 - (\nabla \sigma)^2 - M^2 \sigma^2 + 2 \beta \sigma \dot \pi_c   \nn \\ 
&& + \frac{s  \left( 1 - s^2 \right)}{\Omega^2}  \dot \pi_c (\dot \pi_c^2 -  (\nabla \pi_c)^2 ) + \frac{ s \beta}{\Omega^2}  \sigma \left( \dot \pi_c^2 - (\nabla \pi_c)^2 \right)   \nn \\
&& +  \gamma_1 \sigma^2  \dot \pi_c   +  \frac{\gamma_2}{\Omega} \sigma  \dot \pi_c^2  + \frac{\gamma_3}{\Omega^2}  \dot \pi_c^3  - \xi \sigma^3   \bigg] , 
\label{EFT-pi-sigma-2}
\eea
where $\bar \alpha_1(0) = 1$ normalizes the expansion as before, and $1/s^2 = 1 + 2 \bar \alpha_2(0)$, $\gamma_1 = s \bar \alpha_1''(0)$, $\gamma_2 = 2 \bar \alpha_2'(0) s^2 $, and $\gamma_3 = 2 \bar \alpha_3(0) s^3 $, are all dimensionless parameters (here $'$ represents a derivative in terms of $\sigma/\Omega$). We also introduced $\beta =  s \bar \alpha_1'(0) \Omega$ which has mass dimension. In addition, the normalized Goldstone boson $\pi_c (t, \x)$ is given by $\pi_c \equiv  \Omega^2  s^{-1} \, \pi$. The first line contains the quadratic part of the action, which describes the non-interacting (free) part of the theory. The second and third lines contain the allowed cubic interactions of the theory. Again, we emphasize the role of certain non linear relations tying terms of different orders ({\it e.g.} those relations involving $s$ and $\beta$).

\section{Acoustic and optical phonons in time crystals}

In analyzing (\ref{EFT-pi-sigma-2}) we should be careful to not interpret the parameter $s$ as the speed of sound of $\pi_c$-fluctuations and $M$ as the mass of $\sigma$ fluctuations. The reason is that the coupling $\beta$ is mixing the two fields, therefore modifying the way that fields and modes are related~\cite{achucarro11}. To clarify this point let us study the spectrum of the theory by focusing on the quadratic (free) part of (\ref{EFT-pi-sigma-2}). It is direct to see that the linear equations of motion deduced from (\ref{EFT-pi-sigma-2}), in Fourier space, are given by:
\bea
\ddot \pi_c + s^2 \k^2 \pi_c = - \beta \dot \sigma , \\
\ddot \sigma + (M^2 + \k^2) \sigma =  \beta \dot \pi_c .
\eea
If we consider ansatze of the form $\propto e^{+ i \omega t}$, we obtain the following equation determining $\omega$ as a function of the wave-vector: $(\omega^2 - s^2 \k^2 ) (\omega^2 - M^2 - \k^2) = \beta^2 \omega^2$. From here we can already see that $s$ cannot be interpreted as a sound speed and that $M$ is not a mass unless $\beta=0$. The four solutions to this equation are given by
\bea
\omega_{\pm}^2 &=& \frac{1}{2} \bigg[ \Lambda^2 + \k^2 (1 + s^2)  \nn \\
&& \pm  \sqrt{(\Lambda^2 + \k^2 (1 + s^2) )^2 - 4 s^2 \k^2 (M^2 + \k^2)}  \bigg] , \qquad
\eea
where $\Lambda^2 \equiv M^2 + \beta^2 $. From here we can see that in the long wavelength limit $k \to 0$, we obtain $\omega_-^2 = c_s^2 \k^2$ and $\omega_-^2 = \Lambda^2$, where we have defined
\be
c_s^2 = s^2 M^2 / (M^2 + \beta^2) . \label{speed-of-sound}
\ee
We conclude that the spectrum of the theory consists of a gapless Goldstone boson mode $| G , \k \rangle$ with dispersion relation $\omega_G(\k) = \omega_-(\k)$ characterized by a speed of sound $c_s$ given by (\ref{speed-of-sound}), and a gapped mode $| \Lambda , \k \rangle$ with dispersion relation $\omega_\Lambda(\k) = \omega_+(\k)$, and characterized by a mass:
\be
\Lambda = M s / c_s.
\ee 
These results suggest that the gapless modes $| G , \k \rangle$ resemble acoustic phonons, whereas the gapped modes $| \Lambda , \k \rangle$ behave like optical phonons, encountered in conventional crystals. In what follows we reinforce this analogy by analyzing how they interact and decay.

To study their dynamics, we may proceed to quantize the theory by defining creation and annihilation operators $\hat a^{\dag}_G(\k)$, $\hat a_{G}({\bf k})$, $a^{\dag}_\Lambda(\k)$, and $a_{\Lambda}({\bf k})$ satisfying the standard commutation relations $[\hat a_a(\p), \hat a_b^{\dag}(\q)] = (2\pi)^3 \delta_{ab} \delta^{(3)} (\p - \q)$, with $\hat a_a (\k) | \mu \rangle = 0$, where $a,b=\Lambda,G$. Then, the fields $\pi$ and $\sigma$ are 
\bea
\pi (t,\x) = \sum_a \int_{k} \Big[ \frac{ \pi_a (k) }{(2\pi)^{3/2}} \hat a_a(\k) e^{ i ( \k \cdot \x - \omega_a t)} + {\rm h.c.}Ê\Big] , \quad \\
\sigma (t,\x) =\sum_a \int_{k} \Big[ \frac{ \sigma_a (k) }{(2\pi)^{3/2}}  \hat a_a(\k) e^{ i ( \k \cdot \x - \omega_a t)} + {\rm h.c.} \Big] , \quad
\eea
where $\int_{k} \equiv \int d^3 k$. In the previous expressions, $\pi_G (k)$, $\pi_\Lambda (k)$, $\sigma_G (k)$ and $\sigma_\Lambda (k)$ are the amplitudes of the fields in Fock space obtained by imposing canonical commutation relations between the fields and their momenta. These are found to be given by:
\bea
&& \pi_G (k) = \sqrt{\frac{ (\omega_\Lambda^2 -  s^2 \k^2) \omega_G }{ 2 s^2 \k^2 (\omega_\Lambda^2 - \omega_G^2)}} , \label{pi-G} \\
&& \pi_\Lambda (k) = i \sqrt{ \frac{ (s^2 \k^2 - \omega_G^2 ) \omega_\Lambda }{ 2 s^2  \k^2 (\omega_\Lambda^2 - \omega_G^2)}} , \qquad \\
&& \sigma_G (k) = - i \sqrt{ \frac{ (\omega_\Lambda^2 - M^2 - \k^2) \omega_G }{ 2 (M^2 + \k^2) (\omega_\Lambda^2 - \omega_G^2)}} , \\
&& \sigma_\Lambda (k) = \sqrt{ \frac{ (M^2 + \k^2 - \omega_G^2 ) \omega_\Lambda }{ 2 (M^2 + \k^2)  (\omega_\Lambda^2 - \omega_G^2)} } \label{sigma-Lambda} .
\eea
Then, the phonons of wave-vectors $\k$ are described by states $ | G , {\bf k} \rangle \equiv (2 \pi)^{3/2} \sqrt{2 \omega_G} a^{\dag}_G({\bf k}) | \mu \rangle$ and $| \Lambda, {\bf k} \rangle \equiv (2 \pi)^{3/2} \sqrt{2 \omega_\Lambda} a^{\dag}_\Lambda ({\bf k}) | \mu \rangle$.

\section{The long wavelength limit}

Let us briefly digress to verify that the EFT of eq.~(\ref{EFT-pi}) is recovered from (\ref{EFT-pi-sigma-2}) in the long wavelength limit. Given that we have a gapped mode $| \Lambda , \k \rangle$ of mass $\Lambda = M s/c_s$, it makes sense to ask about the form of the effective field theory in the regime $\omega_-(\k) \ll \Lambda$, where gapless modes $| G , \k \rangle$ are the only relevant degrees of freedom in the system~\footnote{This was the subject of ref.~\cite{Castillo:2013sfa} and we refer to that work for a fully detailed explanation on how to derive this EFT.}. In brief, for long wavelengths $\k^2 \ll M^2$, we may safely assume that $\sigma$ fluctuations are small, and therefore neglect the term proportional to $\xi$ in the action (\ref{EFT-pi-sigma-2}). This implies that the equation of motion for $\sigma$ is of the form $\ddot \sigma + (M^2 - \nabla^2 - \gamma_1 \dot \pi_c) \sigma =  \beta \dot \pi_c + O_2(\pi)$, where $O_2(\pi)$ are terms of second order in $\pi$, which cannot be neglected. In the limit $\k^2 \ll M^2$ this equation may be used to write $\sigma  \simeq (M^2 - \gamma_1 \dot \pi_c)^{-1} \left[ \beta \dot \pi_c + O_2(\pi) \right]$. Inserting this expression back into (\ref{EFT-pi-sigma-2}) to get rid of $\sigma$ in terms of $\pi$, we obtain an action of the form~(\ref{EFT-pi}), with $c_s$ given by (\ref{speed-of-sound}), and $\alpha_3 = \bar \alpha_3 + \bar \alpha_2' \bar \alpha_1' \Omega^2/M^2 + (\bar \alpha_1')^2 \bar \alpha_1'' \Omega^4/ 2 M^4$. This result gives us reassurance that the non linear relations connecting terms of different orders in action (\ref{EFT-pi-0}) is not modified by physics at short wavelengths, as it should. On the other hand, for wavelengths $M^2 \ll \k^2 \ll \Lambda^2$ one may obtain an effective field theory for $\pi$ with a very different structure of space-time differential operators~\cite{Castillo:2013sfa}. This is due to the fact that in this regime the scaling of operators start to change as a signal that there is a new degree of freedom (in this case $| \Lambda, k \rangle$) appearing at shorter wavelengths. 

\section{Dynamics of fluctuations}

We may now discuss the dynamics of the two modes $| G , \k \rangle$ and $| \Lambda , \k \rangle$. To simplify this discussion, without loosing any insight on the physics underlying this class of systems, we consider the choice of parameters $s=1$, $\gamma_2 = \gamma_3 = \xi=0$, and $\gamma_1 = \beta^2/2 \Omega^2$, where $\beta \neq 0$ is the coupling responsible for the phenomenology worth discussing. The choice $\gamma_1 = \beta^2/2 \Omega^2$ simplifies a few of the computations and corresponds to the particular case of a two-scalar field canonical model endowed with an $SO(2)$ symmetry (\emph{e.g.} with a Mexican-hat scalar potential). In such a model, the time translation symmetry is broken by a background trajectory, in field space, moving around the symmetry axis of the Mexican hat at a constant angular speed~\cite{Castillo:2013sfa}. This theory has three independent parameters: $\Lambda = \sqrt{M^2 + \beta^2}$, which is the mass of the gapped mode; $\Omega$, which is the scale associated to the symmetry breaking, and $c_s = M/\sqrt{M^2 + \beta^2}$, which is the speed of sound of the Goldstone boson mode.

\begin{figure}
\centering
\includegraphics[scale=0.30]{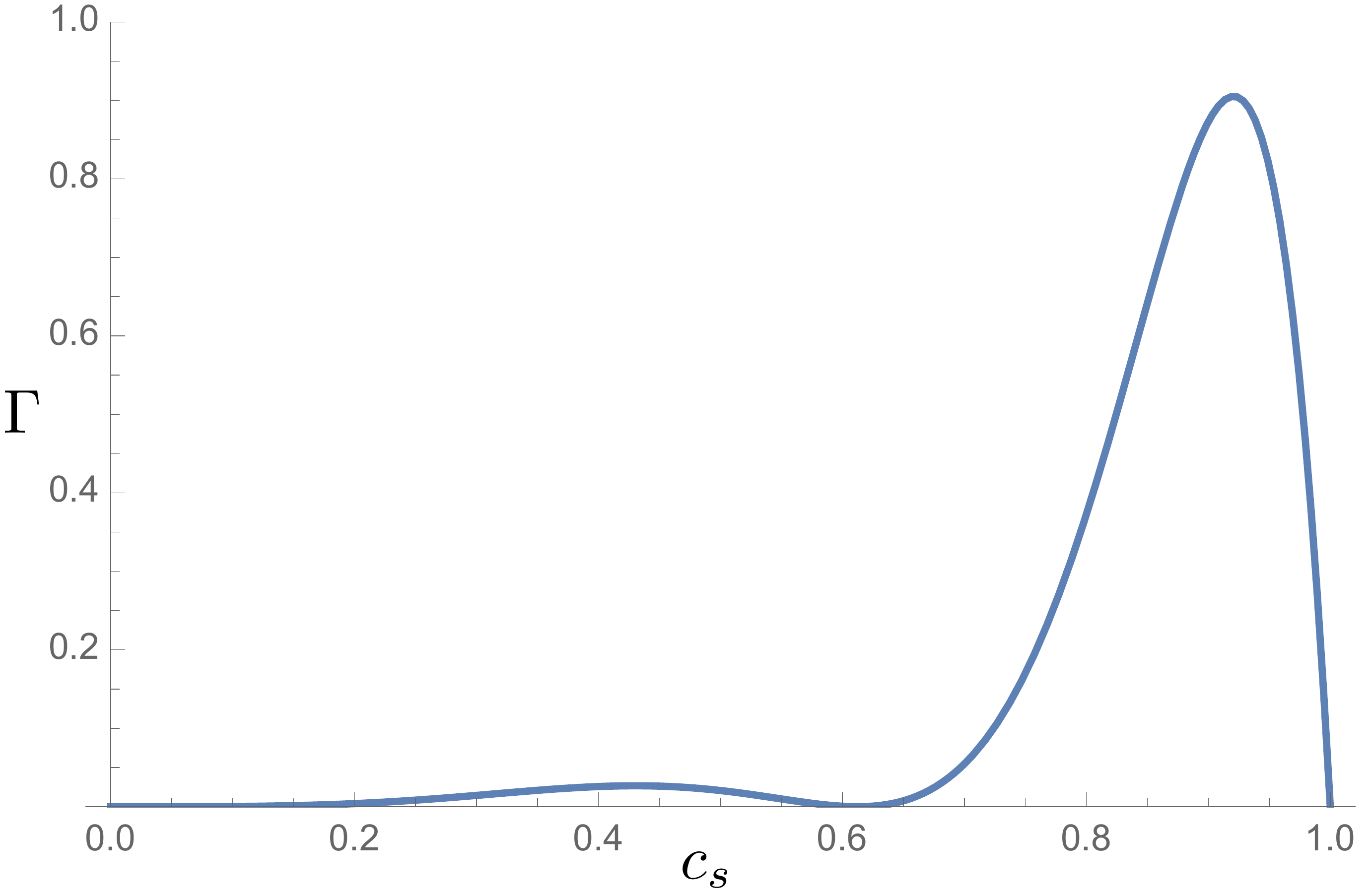}
\caption{The figure shows the decay rate $\Gamma_{\Lambda \to 2 G}$ (in units of $3.5 \times 10^{-4} \Lambda^3 / \Omega^2$) of a $\Lambda$-fluctuation into two Goldstone modes ($\Lambda \to 2 G$) as a function of the speed of sound $c_s$.}
\label{fig:decay-1}
\end{figure}

At lowest order there are two processes that are relevant to understand the dynamics of fluctuations in this class of system. The first one is the generation of a gapped mode $\Lambda$ by the collision of two gapless modes $2G \to \Lambda$, and its subsequent decay.  The second process of interest correspond to the decay of a gapless mode into two gapless modes of shorter wavelengths $G\to 2G$. Both of these processes are due to the same cubic interaction in the $\pi$-$\sigma$ action of eq.~(\ref{EFT-pi-sigma-2}). However, instead of using this form of the action to compute the decay rates of interest, it is possible to simplify the computation by introducing new fields $\tilde \pi$ and $\tilde \sigma$ through a field redefinition~\cite{Castillo:2013sfa} of the form $\pi_c = \tilde \pi (1 - \beta \tilde \sigma / 2 \Omega^2)$ and $\sigma = \tilde \sigma + \beta \tilde \pi^2 / 4 \Omega^2$. This field redefinition does not affect the form of the free field theory, ensuring the validity of (\ref{pi-G})-(\ref{sigma-Lambda}) as the amplitudes for the new fields $\tilde \pi$ and $\tilde \sigma$. On the other hand, the new cubic interaction term in this version of the theory becomes:
\be
S_{\rm int} = - \frac{\Lambda^3 }{4 \Omega^2} c_s^3 (c_s^{-2}-1)^{1/2} \int_x  \tilde \pi^2 \tilde \sigma. \label{Vint}
\ee
With this form of the cubic interaction the computation of these decay rates is straightforward. For instance, at tree level the computation of the $\mathcal M$ matrix for the decay $\Lambda \to 2G$ is deduced from $(2 \pi)^4  \delta^{(3)}(\p_\Lambda - \p_1 - \p_2) \delta(\omega_\Lambda - \omega_1 - \omega_2)  \mathcal M_{\Lambda \to 2G} = - \langle G_1, \p_1 ; G_2, \p_2  | S_{\rm int} | \Lambda , \p_\Lambda \rangle$, and is found to be
\bea
 \mathcal M_{\Lambda \to 2G} = - i \frac{\Lambda^3}{\Omega^2} c_s^3 (c_s^2 - 1)^{1/2} \sqrt{2 \omega_\Lambda (\p) \omega_{G}(\p_1) \omega_{G}(\p_2) }  \nn \\ 
 \times \left[ \sigma_\Lambda  \pi_{G_1}^*  \pi_{G_2}^* + \sigma_{G_1}^* \pi_{\Lambda}  \pi_{G_2}^*  + \sigma_{G_2}^*  \pi_{G_1}^*  \pi_{\Lambda} \right] , \,\, \quad
\eea
where $\pi_G$, $\pi_\Lambda$, $\sigma_{G}$ and $\sigma_\Lambda$ are the amplitudes of eqs.~(\ref{pi-G})-(\ref{sigma-Lambda}).
The desired decay rates are then computed with the help of the Fermi's golden rule.
FIG.~\ref{fig:decay-1} shows the resulting decay rate $\Gamma_{\Lambda \to 2G}$ for $\p = 0$ as a function of $c_s$, keeping both  $\Lambda$ and $\Omega$ fixed. It may be appreciated that the decay rate becomes zero for the value $c_s = \sqrt{3/8}$. This may be traced back to the fact that the fields $\tilde \pi$ and $\tilde \sigma$ appearing in (\ref{Vint}) are linear combinations of the two particle states $\Lambda$ and $G$ with amplitudes as given in (\ref{pi-G})-(\ref{sigma-Lambda}). This linear combination allows the appearance of roots in $\mathcal M$ as one varies the value of $c_s$. The fact that the decay rate becomes suppressed at the value $c_s = \sqrt{3/8}$ tells us that the $\Lambda$-resonance will decay through other channels, such as $\Lambda \to 3 G$, also allowed by the interactions, suppressed by the ratio $M / \Omega$. 

\begin{figure}
\centering
\includegraphics[scale=0.30]{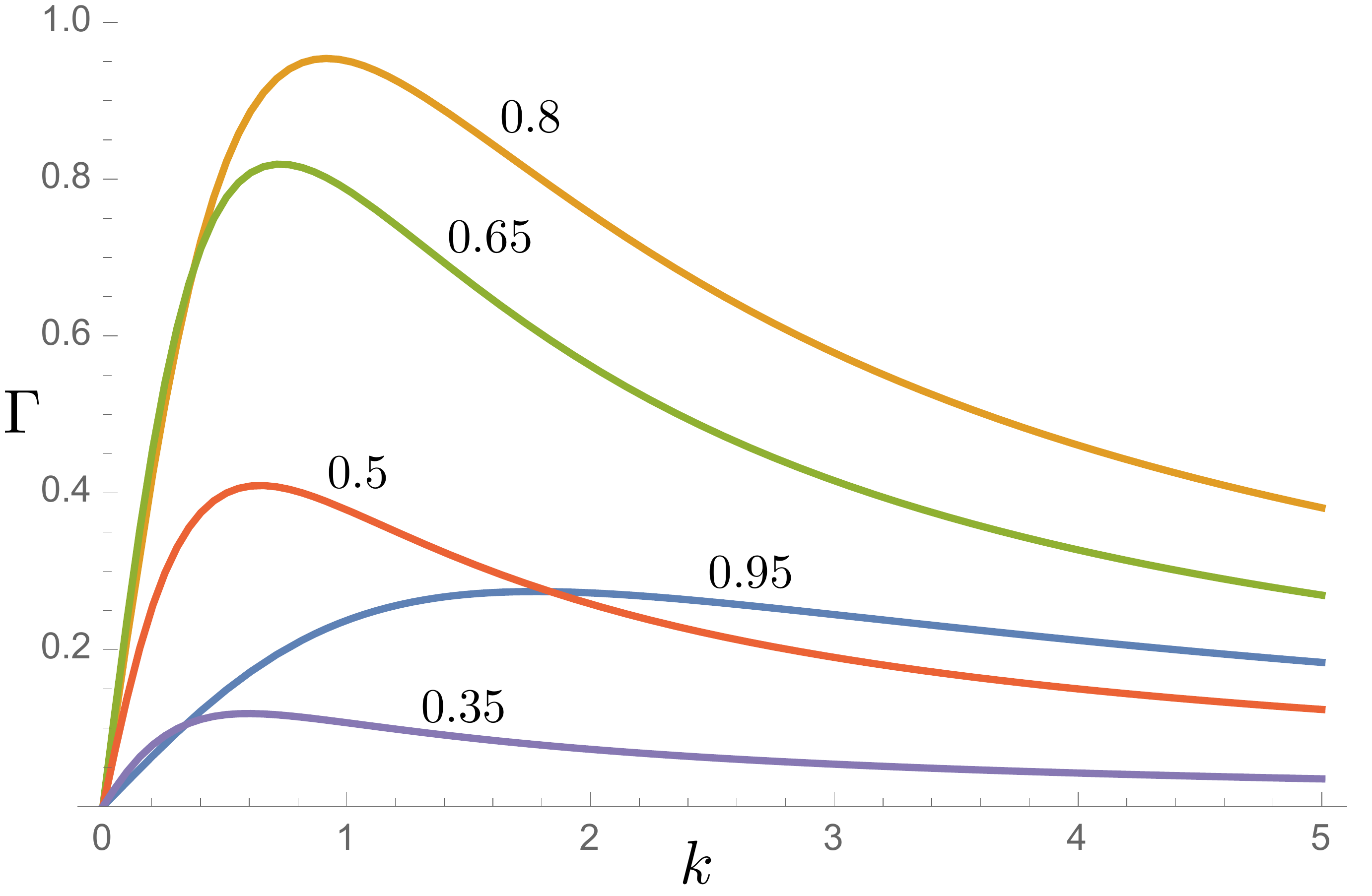}
\caption{The figure shows the decay rate $\Gamma_{G \to 2 G}$ (in units of $4 \times 10^{-5} \Lambda^3 / \Omega^2$) of a Goldstone mode fluctuation into two Goldstone modes ($G \to 2 G$) as a function of the wavelength $k$ of the initial fluctuation $G$. The figure includes the decay rate of the cases $c_s = 0.35$, $c_s = 0.5$, $c_s = 0.65$, $c_s = 0.8$ and $c_s = 0.95$. }
\label{fig:decay-2}
\end{figure}

One may also compute the decay rate $\Gamma_{G \to 2 G}$ for the process $G \to 2 G$. This is plotted in FIG.~\ref{fig:decay-2} as a function of the initial momentum $k= |\k|$ of the decaying Goldstone boson, for several values of $c_s$, keeping both $\Lambda$ and $\Omega$ fixed. 
We would like to stress that fixing $\Lambda$ and $\Omega$ while varying $c_s$ constitutes an arbitrary choice. More generally, these parameters could depend on $c_s$, with a dependence determined by the specific model under consideration. Moreover, this dependence may affect the length scale at which the EFT becomes strongly coupled.

These results show that the Goldstone boson modes $G$ and gapped modes $\Lambda$ are indeed analogous to acoustic and optical phonons respectively, with interesting properties on their own. For instance, we see that $\Lambda$ states are resonances that may be produced through the collision of two $G$-states. This production will be forbidden for certain values of $c_s$ (in the present case $c_s = \sqrt{3/8}$). On the other hand, we see that $G$-states decay into states of longer and longer wavelengths, becoming part of the crystal condensate. These two aspects of time crystals appear to be rather general, and may be central to the phenomenology of this class of systems.

\section{Conclusions}

We conclude by summarizing the main results of this work. First, we have shown that time crystals may be described with an action involving gapless Goldstone boson modes parametrizing the breaking of time translation invariance. The effective field theory for these Goldstone boson excitations is characterized by a non-linear relation which ties together field operators at different orders in perturbation theory. We have found that these Goldstone boson modes may be identified with acoustic phonons found in conventional crystals. Any other gapped fluctuation appearing in the effective field theory can be identified with optical phonons, and may interact with the gapless Goldstone bosons. We have further found that the gapped modes $\Lambda$ are allowed to decay into pairs of Goldstone bosons $G$, and that a single Goldstone boson $G$ is also unstable, and may decay into two Goldstone bosons $G$ characterized by longer wavelengths, which is possible because of their non-relativistic dispersion relations (resulting from the breaking of Lorentz invariance). In the particular example under consideration, the decay rate of the latter process is suppressed by the energy carried by the decaying Goldstone boson, so its mean life time increase as the wavelength becomes larger (recall Fig.~\ref{fig:decay-2}). As a consequence, we find that if the initial state of a time crystal consists of several excitations representing gapless $G$-states, these will inevitably decay into other gapless modes, each time of longer and longer wavelengths (and therefore of less energy). Because the process is suppressed by the energy carried by the gapless modes, these decays is found to be less common with time. The end result is that the state of the system asymptotes to a condensed state inhabited only by gapless modes of very long wavelengths.\\

\section*{Acknowledgements}
We wish to thank Alvaro Nu\~nez and Mar\'ia Jos\'e Santander for useful discussions. 
This work was supported by the Fondecyt projects 1130777
(GAP), 1120360 (BK), the Anillo projects ACT1122 (GAP \& EC), ACT10201 (BK \& EC),
the CONICYT-PCHA/MagisterNacional/2012 - 22120366 scholarship (EC) and the CONICYT/ALMA
31100003 project (EC). 
%

%\vspace{-20pt}

\bibliography{ckp_2}

\end{document}